\documentclass[twocolumn,aps,prb,superscriptaddress,amsmath,floatfix,showpacs]{revtex4}
\usepackage{graphicx}  
\usepackage{subfigure}
\usepackage{amssymb}

\begin{document}

\title{First order phase transitions in Ferromagnet/Superconductor layered structures}
\author{Paul H. Barsic}
\email{barsic@physics.umn.edu}
\affiliation{School of Physics and Astronomy
and Minnesota Supercomputer Institute, University of Minnesota, Minneapolis, Minnesota 55455}
\author{Oriol T. Valls}
\email{otvalls@umn.edu}
\affiliation{School of Physics and Astronomy, University of Minnesota, Minneapolis, Minnesota 55455}
\author{Klaus Halterman}
\email{klaus.halterman@navy.mil}
\affiliation{Physics and Computational Sciences, Research and Engineering Sciences Department, Naval Air Warfare Center,
China Lake, California 93555}

\date{\today}



\begin{abstract}
We study the thermodynamics of clean
structures composed
of superconductor (S) and ferromagnet (F) layers and consisting of 
one or more  SFS junctions. We use 
fully self consistent numerical
methods to compute the condensation
free energies of the possible order parameter configurations
as a function of temperature $T$. As
$T$ varies, we find that there are
phase transitions between states characterized
by different junction configurations (denoted
as ``$0$'' or ``$\pi$''
according to the phase difference of the order parameter in
consecutive S layers). We show that these transitions
are of first order. 
We calculate
the associated latent heats  and find them to
be measurable. 

\end{abstract}


\maketitle




A plethora of new ideas and devices has been emerging from the study of nanostructures
as they pertain to the field of spintronics\cite{iz}. An important part
of this development has occurred through the study of the rich
and varied phenomena that occur\cite{brev} in heterostructures
involving superconductors (S) and ferromagnets (F). 

The physics  of such F/S heterostructures is dominated
by the proximity effects that arise from the competition
between superconducting
and magnetic orderings
in the materials comprising the structure, with each of the corresponding
order parameters penetrating into the other
material. These effects follow from normal and
Andreev \cite{andreev} reflection processes at
the interfaces. In the latter process, an electron
encounters one of the interfaces, is converted into a hole with opposite spin and traverses
in the opposite direction. For S/N interfaces (where N is a non-magnetic,
non-superconducting metal)
the dynamics of charge transport
is degenerate with respect to the electron spin
quantum variable. 
This is not true, however, in the case we consider here, where
superconducting regions are separated by magnetic interlayers.

In SFS trilayers, as 
well as in multilayers built from a sequence of such 
structures (SFSFS\ldots),
the spin-splitting effect of the magnet produces important
and nontrivial changes in 
Andreev and other scattering processes.
The exchange field in the ferromagnet breaks the
time-reversal symmetry and generates a superconducting state
where the Cooper pairs acquire a finite momentum 
resulting in a spatial modulation of the superconducting 
pair amplitude in the F region \cite{demler}. Depending
on the geometric and material characteristics of the SFS trilayer, its
thermodynamic
equilibrium state  can be a ``0'' or a
``$\pi$'' state, depending on the value of the phase difference between
the superconducting order parameters in the two S electrodes. It is this
twofold possibility that lies at the foundation of the many spin-based
switching phenomena, which in turn are the basis for devices,
including  superconducting $\pi$ qubits\cite{yama} and memory 
elements\cite{ioffe}.  
For larger SFSFS\ldots S type heterostructures
the order parameter may or may not flip between
any pair of consecutive S layers, leading to a 
variety of possible configurations, which can be
characterized as a sequence of $0$  and $\pi$ junctions. 

Continual advances in nanoprocessing methods have 
made it possible to fabricate high quality structures containing
SFS junctions, which have
encouraged
further study of these systems.
From the
thermodynamics point of view, 
a $0$ to $\pi$ transition as the temperature 
was varied was inferred\cite{rya,sellier,frolov}  from 
Josephson current measurements in Nb/CuNi/Nb junctions.
The Josephson coupling in similar
structures was also found to cross over
from positive to negative, depending on the F layer thickness\cite{kontos},
indicating again a $0$ to $\pi$ switch.
Under many experimental conditions, a change
in the second-order Josephson 
coupling component
was associated\cite{rya,sellier,kontos} with the $0-\pi$ crossover
while for other conditions\cite{frolov}
the nonlinear current-phase relation did not reveal any change
in the second-order Josephson coupling.

The 
thermodynamics of F/S structures has been directly examined
largely in terms of studying
the transition temperature
to the normal state\cite{buzdin,milo,jin,tollis,pokrovsky}.
A first order phase transition to the
normal state has been predicted\cite{buzdin}
in spin-valve ferromagnet-superconductor-ferromagnet (FSF) nanostructures 
with parallel 
magnetization alignments in thin F layers.
For antiparallel  orientation of the magnetization, 
the transition was found to be always second order\cite{tollis}.
The application of a bias voltage\cite{jin} up to a critical value 
can cause the phase transition to be first order
for both P and AP alignments of the magnetization.
For  F/S bilayers with stripe domain structure\cite{pokrovsky}, 
superconductivity appears via a first order transition.
In some cases, with the action of an applied field, the 
compensating screening currents can enhance\cite{milo} the superconducting 
state. 
The theoretical literature that examines the thermodynamic transitions between
$0$ and $\pi$ configurations in SFS type structures is  more sparse. 
It has been argued\cite{buz}
that  the transition that takes place in 
Josephson junctions is continuous 
when the F layer thickness is uniform but it rounds
off when it is variable.
To further understand the underlying competition between the various
possible states and to better tailor these structures for
practical applications, it is imperative to clarify the thermodynamics of
systems
potentially containing $\pi$ junctions by investigating the parameters that 
may influence
a first or second order phase transition from a 
$0$ state to a $\pi$ state or vice versa.

The objective of this paper is to clarify some of these issues by 
rigorously considering
the thermodynamics of clean layered systems consisting of
one or more SFS junctions so as to identify  and characterize  
any  phase transitions involving $0-\pi$ flipping.
It has been shown at low-temperatures\cite{hv3} that  for 
given S and F widths, exchange energy and other material parameters, multiple
spatial configurations of the self consistent pair amplitude can exist as local
minima of the energy. 
The larger the number of layers, the more combinations were found
to be possible.
Among the various
solutions,  the ground state was found from accurate
condensation energy computations.
Here on the other hand, we have the more ambitious objective
of studying the possible competing states as a function of temperature $T$
through a careful analysis of the free energy differences.
We  find that, as $T$ varies, phase transitions associated 
with flipping of SFS junctions from a $0$ to  a $\pi$
state occur
and that there is a discontinuity in the entropy at such transitions, which
therefore are of first order.  We calculate the
corresponding latent heat, and find that its magnitude
is observable given current experimental capabilities. 

We  
study  layered S/F systems containing
a number $N_J$ of SFS junctions. We consider in particular $N_J=1$, the 
important
case of a single junction, and, to show the richness and variety of the
possible outcomes, also the case $N_J=3$. We 
compare  as a function of
$T$  the condensation
free energies $\Delta {\cal F}(T)$, 
defined as the difference 
$\Delta {\cal F}(T) \equiv {\cal F}_S(T)-{\cal F}_N(T)$
between the free energies of the normal and superconductor states
of each of the several possible self-consistent
configurations.   
It is in principle very difficult to compute
condensation free energies to the
required accuracy: one needs only to recall 
that in the bulk and at zero $T$, $\Delta \cal F$ equals\cite{tink} 
$-(1/2)N(0)\Delta_0^2$ ($\Delta_0$ is
the bulk gap). Since ${\cal F}_N$ is much larger,  
of order $N(0)\omega_D^2$ ($\omega_D$ is the Debye frequency), 
$\Delta \cal F$ is the small difference between two
much larger quantities. 
Thus,  one is faced 
with the tough numerical task of computing two quantities accurately enough
so that the small difference between them can be reliably established.
The problem is exacerbated for our systems 
because, as will be seen, the  difference
between the condensation free energies
of the various possible states (e.g., the 0 and $\pi$ cases for SFS
trilayers) 
is at best
a small fraction of their
individual condensation free energies.
The numerical methods we use do  resolve these small
differences. These calculations
are carried out using the spectra and
pair potentials obtained by fully self
consistent numerical solution of the Bogoliubov deGennes\cite{bdg} (BdG)
equations.

The systems studied here consist then of either 3 or 7 layers,
containing 1 or 3 SFS junctions respectively. 
We denote the thickness
of the F layers by $d_F$ and that
of the  S layers by $d_S$ and assume the interfaces are sharp.
The procedures that we use to numerically solve the BdG equations
for a clean system
in a fully self consistent manner are detailed in Refs.~\onlinecite{hv1,hv3}
and the details need not be repeated here.
As explained there, for many relevant values of the geometric and material
parameters several self consistent solutions, that is, local minima 
of the free energy, often coexist
at $T=0$. For example, solutions\cite{hv3}
of both the $0$ and $\pi$ type may be possible for one junction, while
for the three junction case more than one of the possible symmetric
states ($000$, $\pi\pi\pi$, $\pi 0 \pi$ and $0\pi 0$ in the
obvious notation denoting the state of each junction) may yield 
a self-consistent solution to the problem. In such cases,
the equilibrium  state had to be found by comparing the respective 
energies. 
It was shown that the symmetry of the 
stable solution at $T=0$ can  change\cite{hv3}
as one varies parameters such
as  $d_F$, the interface barrier height, the exchange field
of the magnet, or the Fermi wavevector
mismatch between F and S materials. 
The switchover from one stable state to another occurs as
the 
corresponding  energies cross
at certain points
in parameter space: ``phase transitions'' take place as a function of such
parameters. 
They may occur also in the dirty limit\cite{buz} as a function of $d_F$.
The issue is
how these phenomena are related to  transitions
found experimentally\cite{rya,sellier,frolov,kontos} at finite $T$.
 
\begin{figure}
\centering
\includegraphics[width=3in]{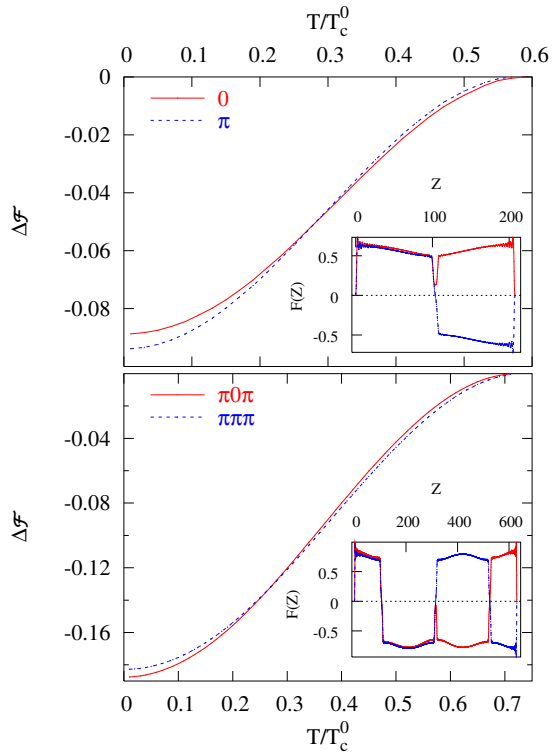}
\caption{(Color online) Normalized condensation free energy $\Delta {\cal F}$
(see text) as 
a function of 
reduced temperature.  The results in the
top panel are for  a one junction (SFS) system  at
$\Lambda=0.535$ and $k_{FS} d_F= 7.5$. Results for the two
possible order parameter configurations are plotted as indicated. Those in the 
bottom panel are for a three junction system at $\Lambda=0.45$
and $k_{FS}d_F=10$. Only the two lowest competing free energy
configurations are plotted. The insets display
the 
normalized pair amplitude $F(Z)$ at the transition points,
where 
$Z\equiv k_{FS}z$.}
\label{fig1} 
\end{figure}

Here we investigate the true thermodynamics, 
that is, 
how
the equilibrium state of a given
system depends on $T$, and  the nature
of the corresponding phase transitions. When self-consistent
numerical solutions of the BdG equations
are possible for multiple order parameter configurations, we   
evaluate 
the condensation free energies of the competing
states. For this purpose we use the method of Ref.~\onlinecite{kos}, which 
requires only  the energy spectrum and 
the position 
dependent order parameter, but not  the eigenfunctions. We find that in many
cases there are first order phase transitions as a function of $T$, 
and that the latent heats are quite appreciable.

To locate phase transitions as a function of $T$,
we  searched near regions of parameter space where
crossings occur\cite{hv3} at zero $T$. 
We 
focussed on the
Fermi wavevector mismatch parameter defined as $\Lambda \equiv E_{FM}/E_{FS}$, 
where\cite{hv1} $E_{FS}$  is the Fermi energy (bandwidth) in 
the S material, while in  F we have
$E_\pm \equiv E_{FM}
\pm h$ for the majority ($+$) and minority ($-$) bands. 
Here $h$ is the exchange field. We also considered the 
geometric parameter, $d_F$.
For $N_J=1$ we take $d_S$
equal to the zero temperature coherence length,
$\xi_0$, while for
$N_J=3$ we double the thickness of the two inner S layers
to $2\xi_0$. This 
mimics a more  symmetric situation, since the outer layers are in contact
with only one F layer. 
We measure energies
in units of $E_{FS}$, lengths in units of $k_{FS}$, the Fermi
wavevector for S, and set $k_{FS} \xi_0= 100$.
The strength of the magnet is 
chosen as 
$h=0.2 E_{FM}$ throughout. We assume 
that all magnetic layers are aligned and that there is no oxide
interlayer barrier. 
We performed several checks of our numerical methods. 
For a system consisting of only one 
S layer with $d_S \gg \xi_0$, we 
quantitatively recover the textbook results\cite{tink} for the
thermodynamics, including the second order transition
at the correct bulk value $T_c^0$ and associated
specific heat discontinuity.

\begin{figure}
\centering
\includegraphics[width=3in]{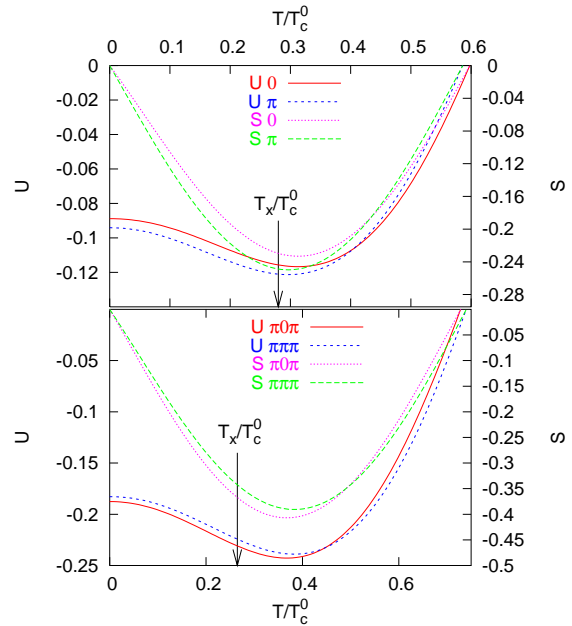}
\caption{(Color online) Dimensionless  (see text)
condensation entropy $S(T)$ (right scales) and
energy $U(T)$ (left scales).  Top  and bottom panels correspond to
the same cases as 
in Fig.~\ref{fig1}.  The vertical arrows 
locate the first-order transitions. 
}
\label{fig2} 
\end{figure}

Proceeding now to the results for $N_J=1,3$, 
the main plots in 
Fig.~\ref{fig1} show the condensation
free energy $\Delta{\cal F}$, as defined above,
normalized by $N(0) \Delta_0^2$, which is twice the value
of the condensation energy
for  a bulk S material at $T=0$. 
In the top panel, $\Delta {\cal F}$ is plotted versus 
reduced temperature $T/T_c^0$ for both the $0$ and $\pi$ configurations 
of a 1 junction 
system.  
These results correspond to
$k_{FS}d_F=7.5$ and $\Lambda=0.535$;
similar results are found at this thickness for a range of
$\Lambda$.
Points at $0.01$ intervals 
in the horizontal temperature scale are plotted, joined by straight segments. 
The slopes of the $\Delta{\cal F}$ curves  approach zero as
$T \rightarrow 0$, which indicates zero entropy at $T=0$.  The slopes
can also be seen to approach zero as $\Delta {\cal F} \rightarrow 0$: 
thus we find 
that the transition to the normal state is
second order, occurring   at 
$T_c < T_c^0$.  
One can also see in the upper plot that 
the results 
for the $0$ and $\pi$
states cross at $T_x/T_c^0 \approx 0.28$, with the
$\pi$
configuration being the 
equilibrium state below $T_x$ and
the $0$ state above. 
The difference in the slopes at $T_x$ represents a latent heat, discussed
below. Thus we predict that
 there is a
transition and that it is first order.
The inset displays the spatial profile of the corresponding pair amplitudes
at $T_x$. The quantity plotted,
$F(Z)$, is the pair amplitude
as usually defined\cite{bdg}, normalized to
its bulk S value at $T=0$. $Z$ is the dimensionless
distance $Z\equiv k_{FS}z$ normal to the layers. 
A discontinuity in the absolute value of $F(Z)$ at the transition can 
be noted. 

The lower panel of Fig.~\ref{fig1} shows  $\Delta{\cal F}(T)$ similarly
plotted for
a 3 junction (7 layer) system with $k_{FS}d_F=10$
and  $\Lambda=0.45$. 
Points  are again taken at intervals of 0.01 on the horizontal scale.
In this case only two 
($\pi\pi\pi$ and $\pi 0 \pi$)
of the possible configuration states 
compete\cite{hv3} as candidates for lowest free energy
over the range of $T$ studied.
The other possible states are therefore 
irrelevant and  omitted.  The  curves in the lower panel
display the same characteristics as in the 1 junction  case: zero first 
derivatives as $T \rightarrow 0$, a second order transition
to the normal state,
and a first order transition at an
intermediate $T$.
In this case the transition is at  $T_x/T_c^0 \approx 0.27$ and from a
$\pi0\pi$ state at lower $T$ to 
$\pi\pi\pi$  at higher $T$. The inset again shows
the corresponding $F(Z)$ at the transition. Thus in this more complicated
example, the
transition involves the inner $0$ junction  flipping to $\pi$
as $T$ is increased. Again, such first-order transitions
exist over a range of $\Lambda$ at fixed $d_F$.


We can now  derive the entire thermodynamics. Thus,
Fig.~\ref{fig2} shows the dimensionless condensation
entropies, $S(T)$, obtained by differentiating
the results for $\Delta {\cal F}(T)$  (Fig.~\ref{fig1}) with respect to 
$T/T_c^0$.  We show also the corresponding condensation
energies, $U(T)$, computed
from standard thermodynamic relations, 
normalized in the same way as $\Delta{\cal F}$.
In calculating the derivatives, a 
polynomial form was fit to the $\Delta{\cal F}$ data in Fig.~\ref{fig1}. 
The values found were equivalent to those
obtained
by taking a discrete derivative of the $\Delta{\cal F}$ curves and fitting
a polynomial to the resultant data.
The top and bottom panels of Fig.~\ref{fig2} 
correspond, respectively, to the
1 and 3 junction
cases  in Fig.~\ref{fig1}.
As mentioned previously,
the entropies go to zero 
smoothly as $T \rightarrow 0$ and $T \rightarrow T_c$. On the other hand,
it is
quite obvious that the transitions at $T_x$ 
(vertical arrows) are indeed first order: 
the entropies of the two states involved  are
not equal at that temperature.
The condensation energy curves resemble those for a bulk superconductor. 
By taking a further derivative,
the specific heat can also be obtained. For any given state the 
quantities $S(T)$, $U(T)$ and $\Delta{\cal F}$ all
go to zero at the same temperature, which is 
the computed value of $T_c$.  The energies and entropies 
cross at temperatures above $T_x$: one can see
that both entropy and energy play important and subtle roles in
the  first-order  transition. 

\begin{figure}
\centering
\includegraphics[width=3in,angle=0]{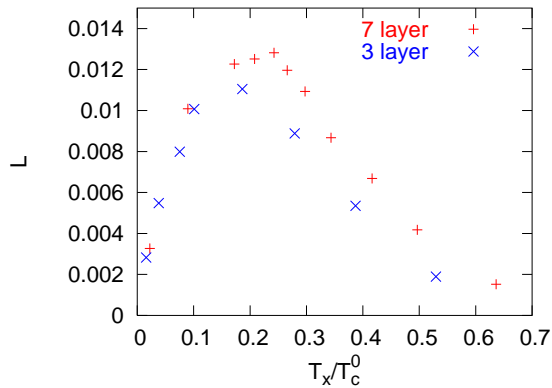}
\caption{(Color online) Variation
of the  latent heat
with $T_x$. The  entropy change, $L$, at $T_x$, normalized to the
normal state specific heat at $T_c^0$, is plotted as a function
of $T_x/T_c^0$ for the one and three junction cases.
}
\label{fig3} 
\end{figure}

As one varies $\Lambda$,  $T_x$ moves up or down
smoothly and monotonically until
the transition disappears when $T_x$ reaches either 0 or $T_c$. 
The difference between the entropies at $T=T_x$, 
which measures the
latent heat, $L$, is quite appreciable and  does not depend 
drastically on $T_x$ (nor equivalently on $\Lambda$) over most
of the range.
These points are illustrated in Fig.~\ref{fig3}, where $L$ 
is shown for both the 1 and 3 junction systems.
We show $L$ (calculated as discussed 
in connection with Fig.~\ref{fig2}) at several $T_x$,
obtained
by varying $\Lambda$. The sign of $L$ is defined by 
subtracting the entropy of the stable
state below $T_x$ from that above $T_x$. So that the size of the effect can
be appreciated, $L$ is  normalized to the
specific heat at $T_c^0$ of  a normal S material of the same volume.
This is appropriate since for a normal metal the specific heat
at $T_c^0$ equals its entropy.
One can see that, for both one and three junction
samples, the entropy jumps at $T_x$
can exceed $1\%$
of the entropy present in a uniform bulk S sample at $T_c^0$. 


Translating these results for the normalized measure of the latent heat into
actual units for typical samples\cite{rya}, we find that they are of order
of up to $10^{10}  k_BT$ or $10^{-13} J$. Standard ac calorimetry techniques
offer a resolution\cite{bou,cc} at least one order of magnitude smaller. 
Indeed, even attojoule
calorimetry has been achieved in electronic systems, 
although by using multiple samples.\cite{bou}
Our rigorous results are clearly larger than estimates
made\cite{buz} for dirty systems. These estimates, based on partial
consideration 
of  the Josephson energy only, are in the
nature of lower bounds.

In summary, we have rigorously studied
the thermodynamics of clean single and triple SFS
junction systems. We have shown
that  as $T$ is varied a given junction can flip from the $0$ state to the
$\pi$ state. The resulting
phase transition is first-order, in agreement
with the experiments of Refs.~\onlinecite{rya,sellier,kontos},
for systems with sharp interfaces. 
The associated latent heats were found to be
within current experimental resolution.
The results presented here should be 
applicable over a broad range of material parameters
assuming
$d_S$ does not exceed several $\xi_0$.

This work was supported in part by the University
of Minnesota Graduate School and by 
the ARSC at the University of 
Alaska Fairbanks (part of the DoD HPCM
program).

\end{document}